\documentclass[aps,prx,reprint,superscriptaddress,nofootinbib,preprintnumbers]{revtex4-2}

\usepackage{amsmath,amssymb}
\usepackage{array}
\usepackage{booktabs}
\usepackage{graphicx}
\usepackage{url}
\usepackage{xcolor}
\usepackage[colorlinks=true,linkcolor=blue,citecolor=blue,urlcolor=blue]{hyperref}
\usepackage{natbib}

\graphicspath{{figures/}}

\newcommand{\rch}{\ensuremath{R_{\rm ch}}}

\newcommand{\bedown}{\ensuremath{B(E2;2_1^+\!\rightarrow0_1^+)}}
\newcommand{\BEtwo}{\ensuremath{B(E2)}}
\newcommand{\rms}{\ensuremath{\mathrm{RMS}}}

\definecolor{nuclrorange}{HTML}{D95F02}
\definecolor{stlblue}{HTML}{4C78A8}
\definecolor{dchgreen}{HTML}{66BFA5}
\definecolor{bskgpurple}{HTML}{9296C9}
\definecolor{modelgray}{HTML}{7A7A7A}
\newcommand{\rmsbar}[2]{\makebox[2.20cm][l]{\textcolor{#1}{\rule{#2}{4.2pt}}}}

\begin{document}

\preprint{CERN-TH-2026-202}

\title{Learning Nuclear Structure with AI: Radii and Collectivity}
\author{Giuliano Giacalone}
\affiliation{Theoretical Physics Department, CERN, CH-1211 Geneva 23, Switzerland}
\author{Sokratis Trifinopoulos}
\affiliation{Theoretical Physics Department, CERN, CH-1211 Geneva 23, Switzerland}
\affiliation{Physik-Institut, University of Zurich, Winterthurerstrasse 190, 8057 Zurich, Switzerland}
\affiliation{Department of Physics and Astronomy, Northwestern University, Evanston, IL 60208, USA}
\author{Mike Williams}
\affiliation{Laboratory for Nuclear Science, Massachusetts Institute of Technology, Cambridge, Massachusetts 02139, USA}
\affiliation{NSF AI Institute for Artificial Intelligence and Fundamental Interactions, Cambridge, Massachusetts 02139, USA}
\date{\today}

\begin{abstract}
Low-energy nuclear structure is encoded in a broad body of experimental
information across the chart of nuclides. Learning how this information is
organized across observables and nuclei can provide a data-driven empirical
baseline for theoretical extrapolations and experimental design. Here, we
develop held-out ensembles based on NuCLR (Nuclear Co-Learned
Representations), a multi-task model of nuclear data, to study charge radii and
electric-quadrupole transition strengths. Out-of-fold (OOF) validation shows that shared representation improves performance over single-task
learning, yielding a charge-radius \rms\ deviation of $0.0147$~fm and a
\BEtwo\ \rms\ deviation of $0.192~e^2{\rm b}^2$ across hundreds of nuclides, competitive with state-of-the-art nuclear models. Our error bars 
estimate the expected prediction accuracy across the nuclear chart, 
highlighting regions where new data would encode information beyond the learned patterns.
NuCLR thus serves as a data-driven surveyor of nuclear
structure and a step toward a shared, multi-observable foundation
model of the nuclear chart.
\end{abstract}

\maketitle

\textit{\textbf{Introduction}}~--~Atomic nuclei have been studied experimentally for more than seven decades, producing a comprehensive body of data on how the many-body dynamics of protons and neutrons shape the building blocks of visible matter \cite{Verney:2025efj}. Patterns in these data have been interpreted and categorized thanks to the use of effective concepts, for example, shell, deformation, and pairing phenomena, which represent the basis of powerful phenomenological approaches, such as self-consistent mean-field (or energy density functional, EDF) theory \cite{Bender:2003jk}. More recently, enabled by advances in scalable many-body numerical methods \cite{Hagen:2013nca,Hergert:2015awm,Frosini:2021fjf,Lee:2025req,Scalesi:2026lja}, focus is shifting toward an \textit{ab initio} derivation of nuclear phenomena \cite{Hergert:2020bxy,Ekstrom:2022yea}. This long-term program aims to describe low-energy nuclear structure as emerging from an effective field theory of the underlying quantum chromodynamics \cite{Epelbaum:2008ga,Machleidt:2011zz,Hammer:2019poc}.

Achieving this goal across the entire nuclear chart is a formidable challenge, and automation provides a key enabler \cite{Frame:2017fah,Duguet:2023wuh,Cook:2024toj,Yu:2025fyh}. Recent developments in emulator-based acceleration and Bayesian methods have opened the possibility of exploring inter-dependencies between observables and model parameters at scale. In the context of \textit{ab initio} computations, emulators now make it possible to evaluate expensive nuclear models rapidly, enabling global sensitivity analyses and uncertainty quantification \cite{Ekstrom:2019lss,Wesolowski:2021cni,Hu:2021trw,Belley:2023lec,Sun:2024iht,Belley:2025nkn,Heihoff:2026ycq,Munoz:2026aem}. 
Ultimately, these advances are essential for identifying where existing descriptions fail to capture patterns in nuclear data and for designing future experiments that would most advance our understanding of nuclear forces and the nuclear many-body problem.

In this Letter, we develop dedicated held-out ensembles based on NuCLR
(Nuclear Co-Learned Representations), a previously introduced task-conditioned
multi-observable model of nuclear data~\cite{Kitouni:2023rct}. 
NuCLR has already learned
a shared representation across binding energies, charge radii, separation
energies, and decay $Q$ values; subsequent analyses showed that its mass
representation contains compact physical organization rather than acting as a
look-up table~\cite{Kitouni:2024ulw,Richardson:2025dze}. 
Here, we use this learned multi-observable
representation to construct an empirical baseline for the evolution of charge
radii and quadrupole transition strengths along isotopic chains. Comparisons
with state-of-the-art nuclear models then reveal where experimentally encoded
patterns in nuclear data are reproduced, while
disagreements in unmeasured regions identify measurements that can discriminate
between empirical and theory-driven extrapolations. In this way, NuCLR acts as
a physics surveyor of the nuclear chart, highlighting regions in which new
measurements would be especially informative.

\textbf{\textit{Target observables}}~--~Charge radius ($R_{\rm ch}$) measurements represent a standard benchmark in the context of EDF approaches, which, as we shall also see here, describe such quantities extremely accurately \cite{Perera:2021ztx}. While a liquid-drop picture of nearly saturated nuclear matter gives the leading
geometric scaling \(R_{\rm ch}\propto A^{1/3}\)~\cite{Myers:1969zz}, deviations from this trend inform us about variations in shell structure, deformation, and surface diffuseness. Isotope-shift-like measurements are especially valuable for challenging theoretical paradigms because they isolate these differential trends with extremely high precision \cite{Campbell:2016yxe,Yang:2022wbl}.

Quadrupole transition strengths probe a different, although related, aspect of nuclear structure. In even-even nuclei, the reduced electric-quadrupole transition probability between the first ($2^+$) state and the ground state represents the standard indicator of collectivity and deformation. Throughout this work,  we write \BEtwo\ for the downward transition strength, \bedown, in units of ($e^2{\rm b}^2$).
In the axial rotor limit \cite{Pritychenko:2013gwa}, one has
\begin{equation}
\label{eq:BE2beta2}
B(E2;2^+ \to 0^+)
=
\frac{9}{80\pi^2}
Z^2 e^2
R_0^4\,
\beta_2^2, 
\end{equation}
where $R_0 = 1.2 \,A^{1/3}$ [fm], and $\beta_2$ is the quadrupole deformation parameter. \BEtwo\ values provide a view of the nuclear chart where nuclei are organized according to the deformation of their shape. These observables are thus highly sensitive to many-body correlation effects in the nuclear states, such that, unlike charge radii, they are not used as a precision benchmark for mean-field approaches. Notably, whereas radii vary smoothly, \BEtwo\ values can change by orders of magnitude along an isotopic chain. 
Therefore, \rch\ and \BEtwo\ test if a learned nuclear representation can capture both slowly varying bulk trends and rapidly changing many-body phenomena.

\textbf{\textit{Objectives and methods}}~--~Existing machine-learning studies related to nuclear structure are essentially of two types. The first are surrogate models and emulators to help interpret and speed up complex many-body computations, and quantify theoretical uncertainties (see e.g.\ \cite{Lasseri:2019ywk,Phillips:2020dmw,Bonilla:2022rph,Drischler:2022ipa,10.3389/fphy.2022.1054524,Verriere:2022llh,Lay:2023boz,Zhang:2024nqr,Lay:2026zbq} for progress within EDF theory). The second category pertains to purely data-driven approaches where relations among observables are derived from experimental data. Previous studies of charge radii \cite{Akkoyun:2012yf,Wu:2020bao,Dong:2022wkd,Cao2023ChargeRadiiCNN} use feed-forward, Bayesian neural networks or single-observable regression. Related approaches have also been applied to \BEtwo\ systematics \cite{Akkoyun2015BE2ANN,Berbache:2025jor}. 

Our model, NuCLR, belongs to the second category of data-driven approaches,
although substantially improving on the state of the art. Rather than training
independent regressors for each observable, NuCLR learns a task-conditioned
representation shared across several nuclear observables~\cite{Kitouni:2023rct}. This
distinction allows information encoded throughout the adopted nuclear-data record to
inform predictions for a target observable, making NuCLR a useful empirical
complement to nuclear theory. For this role, the learned representation should
be $(i)$ coherent, in the sense that new observables are incorporated into a
common representation rather than treated as isolated one-off regression
tasks; $(ii)$ interpretable, at least in regimes where the learned representation
can be related to known nuclear mechanisms~\cite{Kitouni:2024ulw,Richardson:2025dze}; and $(iii)$ robust,
with predictions accompanied by out-of-fold (OOF) validation, quantified
uncertainties, and error diagnostics that expose not only average performance
but also tail behavior. The present work thus tests whether the shared representation learned by NuCLR can infer nuclear size and quadrupole collectivity from the broader experimental record, with the aim of forming the basis for a foundation model of the nuclear chart.

The architecture, fold construction, and complete observable and task list are given in the Supplemental Material.
The main feature of our validation protocol is that it is designed to mimic the situation in which a target value is unknown.
For each observable, the nuclei for which measurements are available are split into ten validation folds.
For a given fold, four independently seeded models omit the same target labels, giving 40 finetuned models per campaign.
When a nucleus for which data are available is scored, only the four models assigned to its held-out fold are used. When the target nucleus has no measurement associated, the full ensemble of 40 models is used. In the charge-radius campaign, one shared auxiliary pretraining stage excludes \rch\ everywhere, after which the radius task is activated while the mass, separation-energy, decay-energy, and identity tasks remain active.
In the \BEtwo\ campaign, ten fold-specific conditional pretrains exclude \BEtwo\ everywhere and also hide same-nucleus quadrupole proxies (such as $E(2^+)$, $E(4^+)$, \ldots) on the active validation fold.

We first construct the ensemble prediction. Let $f_{\theta_m}(Z_i,N_i,q)$ denote the prediction of replica $m$, with network parameters $\theta_m$, for nucleus $i$ and observable $q$.
Here, ${\cal S}_i$ is the set of ensemble members used to score nucleus $i$, and $|{\cal S}_i|$ is the number of models in that set.
For a target with available data, ${\cal S}_i$ contains only models that held out that target. For a target with no experimental information, ${\cal S}_i$ is the full 40-member ensemble.
The central prediction is
\begin{equation}
  \hat y_i =
  \frac{1}{|{\cal S}_i|}\sum_{m\in{\cal S}_i}
  f_{\theta_m}(Z_i,N_i,q).
  \label{eq:ensemble}
\end{equation}
When data are available, the OOF residual is defined as $|y^{\rm exp}_i-\hat y_i|$. For unmeasured nuclei, the interval model estimates the expected residual scale from measured nuclei with similar structural features.

In a second step we use the OOF residuals to determine errors, and establish where NuCLR tends to be more or less accurate. 
To this aim, a separate model is trained to learn how the residual strength varies across nuclei. We use a cross-fitted variant of locally adaptive conformal regression~\cite{Papadopoulos:2002icm,Lei:2018distributionfree}, learning a local error scale from residuals of the four-member OOF predictor for measured nuclei. A separate gradient-boosted regressor \(h_\phi\) takes nuclear-structure and chart-location features \(x_i\) as inputs and predicts the typical magnitude of the OOF residual for nuclei with those features. Exponentiation gives a positive local error scale, \(a_i=\exp[h_\phi(x_i)]\), describing a relative variation across the nuclear chart. The resulting interval half-width (our error bar) is
\begin{equation}
 \Delta_{p,i}=c_p a_i \,,
 \label{eq:interval-main}
\end{equation}
where a disjoint set of measured nuclei is used to determine a common multiplier, \(c_p\), for the desired coverage level, \(p\), while a third disjoint set of measured nuclei is used to evaluate coverage.
The residual-model inputs, gradient-boosting settings, calibration procedure, and complete coverage tests are described in the Supplemental Material. On randomly held-out measured nuclei, the resulting intervals achieve near-nominal coverage for both observables. Random-fold and contiguous-region coverage results are reported separately in Table S1.
The upcoming figures show intervals constructed using the \(p=0.68\) calibration.

Our results are compared to two representative models employed in global-scale 
calculations of nuclear observables within the EDF framework. The first is a Brussels--Skyrme-on-a-grid (BSkG3) EDF model based on self-consistent Skyrme--Hartree--Fock--Bogoliubov (HFB) calculations \cite{Grams:2023sml}. The model tabulates charge radii and ground-state deformations, but not spectroscopic $B(E2)$ transition strengths. Therefore, in our comparison, the BSkG3 values of $B(E2;2_1^+\to0_1^+)$ are obtained from the tabulated quadrupole deformation parameter, $\beta_2$, using the rigid-rotor formula in Eq.~(\ref{eq:BE2beta2}).
The second model is from Delaroche \textit{et al.} \cite{Delaroche:2009fa}. It is based on self-consistent Gogny-D1S HFB calculations which are later mapped onto a five-dimensional collective Hamiltonian (5DCH) to describe ground-state properties and low-lying spectra of even-even nuclei beyond the mean-field approximation. The data tables are published in the AMEDEE database \cite{HilaireGirod_AMEDEE}, and provide both charge radii and spectroscopic observables, including $B(E2;2_1^+\to0_1^+)$.

Before discussing individual case studies, we present in Tab.~\ref{tab:metrics} the global performance of the models considered, as well as a separate multiplicative-error tail diagnostic for the central predictions.
The absolute \rms\ values put NuCLR, 5DCH, and BSkG3 on a similar scale.
One of the most important results of this work is the reference comparison with a single-target-learning (STL) campaign. Multi-task learning (MTL) reduces the radius \rms\ from $0.0760$ to $0.0147$ fm and the \BEtwo\ \rms\ from $0.295$ to $0.192~e^2{\rm b}^2$.
The substantial reduction shows the benefit of sharing information among nuclear observables when inferring charge radii and \BEtwo\ values from the experimental record. This makes NuCLR a useful empirical comparator for nuclear-model predictions.

\begin{table}[t]
\centering
\scriptsize
\setlength{\tabcolsep}{2pt}

\begin{tabular*}{\columnwidth}{@{\extracolsep{\fill}}lrl@{}}
\toprule
Observable/Method & Value & \\
\\\multicolumn{3}{@{}l}{\textbf{Charge radii \rms\ [fm]}}\\
\midrule
Target-only STL OOF & 0.0760 & \rmsbar{stlblue}{2.20cm}\\
\textbf{NuCLR MTL OOF} & \textbf{0.0147} & \rmsbar{nuclrorange}{0.43cm}\\
BSkG3 & 0.0233 & \rmsbar{bskgpurple}{0.68cm}\\
5DCH & 0.0255 & \rmsbar{dchgreen}{0.74cm}\\
\addlinespace

\multicolumn{3}{@{}l}{\textbf{Charge radii: 95th-percentile multiplicative error}}\\
\midrule
\textbf{NuCLR MTL OOF} & \textbf{1.0069} & \rmsbar{nuclrorange}{2.18cm}\\
BSkG3 & 1.0116 & \rmsbar{bskgpurple}{2.19cm}\\
5DCH & 1.0148 & \rmsbar{dchgreen}{2.20cm}\\
\addlinespace

\multicolumn{3}{@{}l}{\textbf{\BEtwo\ \rms\ [$e^2{\rm b}^2$]}}\\
\midrule
Target-only STL OOF & 0.2952 & \rmsbar{stlblue}{2.20cm}\\
\textbf{NuCLR MTL OOF} & \textbf{0.1923} & \rmsbar{nuclrorange}{1.43cm}\\
BSkG3 & 0.1918 & \rmsbar{bskgpurple}{1.43cm}\\
5DCH & 0.1754 & \rmsbar{dchgreen}{1.31cm}\\
\addlinespace

\multicolumn{3}{@{}l}{\textbf{\BEtwo: 95th-percentile multiplicative error}}\\
\midrule
\textbf{NuCLR MTL OOF} & \textbf{3.05} & \rmsbar{nuclrorange}{1.56cm}\\
5DCH & 4.29 & \rmsbar{dchgreen}{2.20cm}\\
\bottomrule
\end{tabular*}

\caption{Validation performance and nuclear-model baselines.
NuCLR MTL and STL entries are OOF predictions on the corresponding measured targets (868 for charge radii and 433 for \BEtwo).
Nuclear-model rows are evaluated where the corresponding baseline is available. For instance, the 5DCH radius overlap contains 337 measured radii, and the \BEtwo\ baseline overlap contains 428 of the 433 measured transitions. The \BEtwo\ multiplicative-error rows use this common 428-transition NuCLR--5DCH overlap.
The 95th-percentile multiplicative error is computed from the symmetric ratio
$\max(y_{\rm pred}/y_{\rm exp},y_{\rm exp}/y_{\rm pred})$.
Thus, $F=3.05$ means that 95\% of predictions differ from experiment by less than a factor 3.05.
It is included because \BEtwo\ spans orders of magnitude, so an absolute \rms\ can look acceptable while missing weak transitions by large factors.
Bars are scaled within each block, and shorter is better.}
\label{tab:metrics}
\end{table}

\begin{figure*}[t]
\centering
\includegraphics[width=\textwidth]{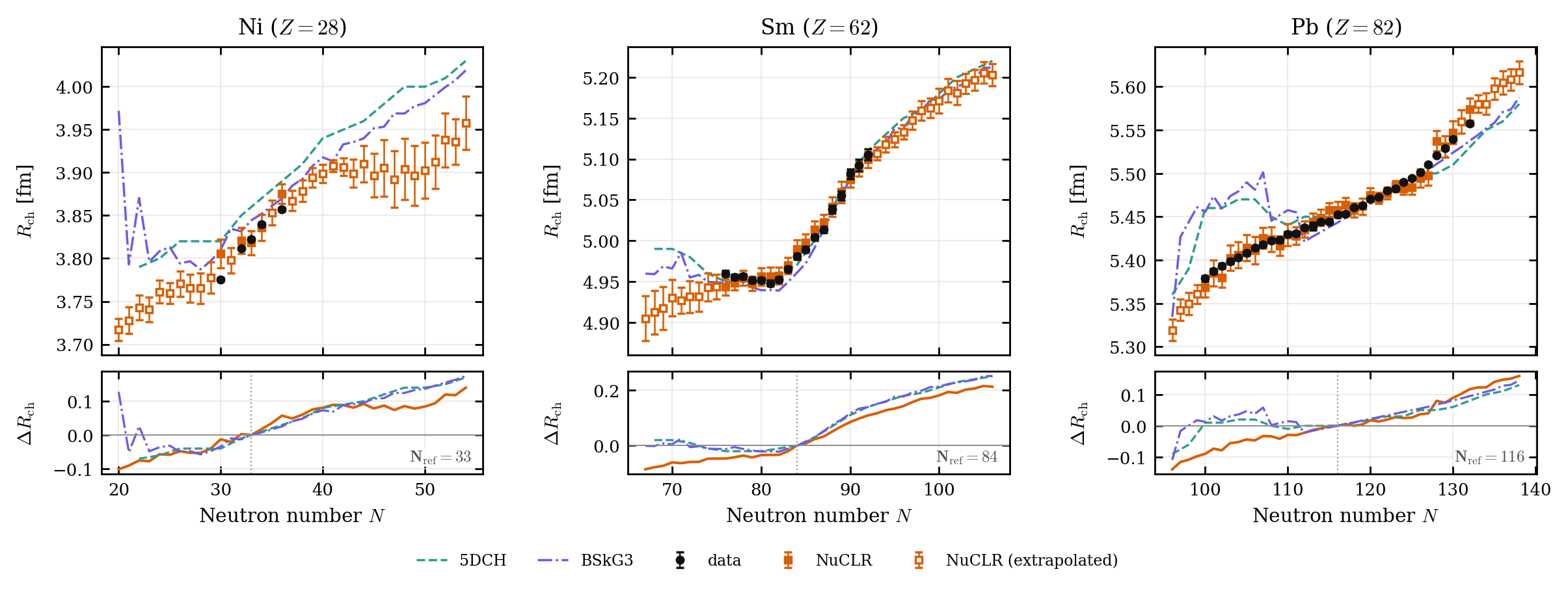}
\caption{
Charge-radius survey for Ni, Sm, and Pb.
Black circles are data; orange squares are NuCLR predictions, filled for held-out measured nuclei and open for extrapolated nuclei.
For filled measured nuclei, orange error bars are cross-fitted 68\% predictive intervals. 
For open extrapolated nuclei, the bars report the OOF residual scale learned from measured nuclei with similar structural features; they are not validated 68\% intervals.
Green dashed and purple dash-dotted curves denote the 5DCH and BSkG3 baselines, respectively.
The axes are zoomed to the data/NuCLR region; baseline segments outside that window are omitted.
Lower subpanels show $\Delta R_{\rm ch}(N)=R_{\rm ch}(N)-R_{\rm ch}(N_{\rm ref})$, with $N_{\rm ref}$ chosen as the central measured isotope of each chain; each model is subtracted by its own value at the same $N_{\rm ref}$.
}
\label{fig:radii-main}
\end{figure*}

\textbf{\textit{Charge radii}}~--~Figure~\ref{fig:radii-main} collects radius chains representative of the capabilities offered by our approach. 

The first is the nickel (Ni) chain, shown in Fig.~\ref{fig:radii-main}(left), whose charge radii have been analyzed extensively to benchmark \textit{ab initio} and EDF calculations \cite{Kaufmann:2020gbf,Pineda:2021shy,Kortelainen:2021raz,Malbrunot-Ettenauer:2021fnr,Sommer:2022sok}. Experimental measurements are rather scarce, but all models including NuCLR are close to the experimental results. 
More interesting are the different continuations predicted on the neutron-rich and neutron-deficient sides.
At large $N$, the chain approaches the doubly-magic $^{78}$Ni \cite{Taniuchi:2019pen}. The NuCLR results predict kinks and a flattening of the charge-radius evolution for $40<N<50$, while the nuclear theory predictions are smooth and show no kinks. Conversely, toward the low-$N$ side, the nuclear models tend to flatten, while NuCLR keeps decreasing. 
The distinct data-driven and theory-driven trends corroborate neutron-rich Ni isotopes as high-impact targets for future charge-radius measurements~\cite{Reilly:2912232}.

The second chain is a rare earth, samarium (Sm), shown in the middle panel of Fig.~\ref{fig:radii-main}. Here, we want to test whether NuCLR captures features associated with the sharp transition from a spherical to a well-deformed nuclear shape, a key characteristic of Sm isotopes occurring around $N=82$. The measured Sm radii display the rapid increase associated with the onset of rare-earth deformation above the $N=82$ shell closure. The OOF NuCLR points track this change of slope, while in the unmeasured neutron-deficient region the data-driven and nuclear-model continuations begin to separate.

Finally, in Fig.~\ref{fig:radii-main}(right) we study the diametrically opposite case of lead (Pb) isotopes, where we move around the doubly-magic point at $Z=82$, $N=126$. Here, the interest is twofold. First, Pb radii are measured across a broad range of $N$, providing an unusually extensive OOF test of whether NuCLR reproduces the isotopic trend when each measured radius is predicted by models that exclude that target.  Secondly, for Pb nuclei below the shell closure, shape mixing and coexistence are known to be important \cite{DeWitte:1056187,Heyde:2011pgw}, which offers an interesting ground for testing our extrapolations. The NuCLR results provide an excellent description of the data in the full range of $N$. The data-driven extrapolation predicts, in particular, a smooth trend toward low values of $N$, in contrast to the visible kinks predicted by the nuclear models.

In the Supplemental Material, results for charge radii in additional representative chains are shown and discussed, respectively, calcium (Ca), tin (Sn), and mercury (Hg). In addition, we show predictions for odd-$Z$ promethium (Pm). Absolute Pm radii are absent from the adopted training table, but published isotope shifts provide changes in mean-square charge radii for $^{143-147}$Pm \cite{Studer:2020hsb}; we compare these independent differential data with NuCLR in Fig.~\ref{fig:promethium}. Measurements farther from stability remain a proposed experimental target \cite{Chrysalidis:2912235}. A global scan across nuclides is provided in the GitHub data repository associated with this work~\cite{NuCLRApplicationsData}.

\begin{figure*}[t]
\centering
\includegraphics[width=\textwidth]{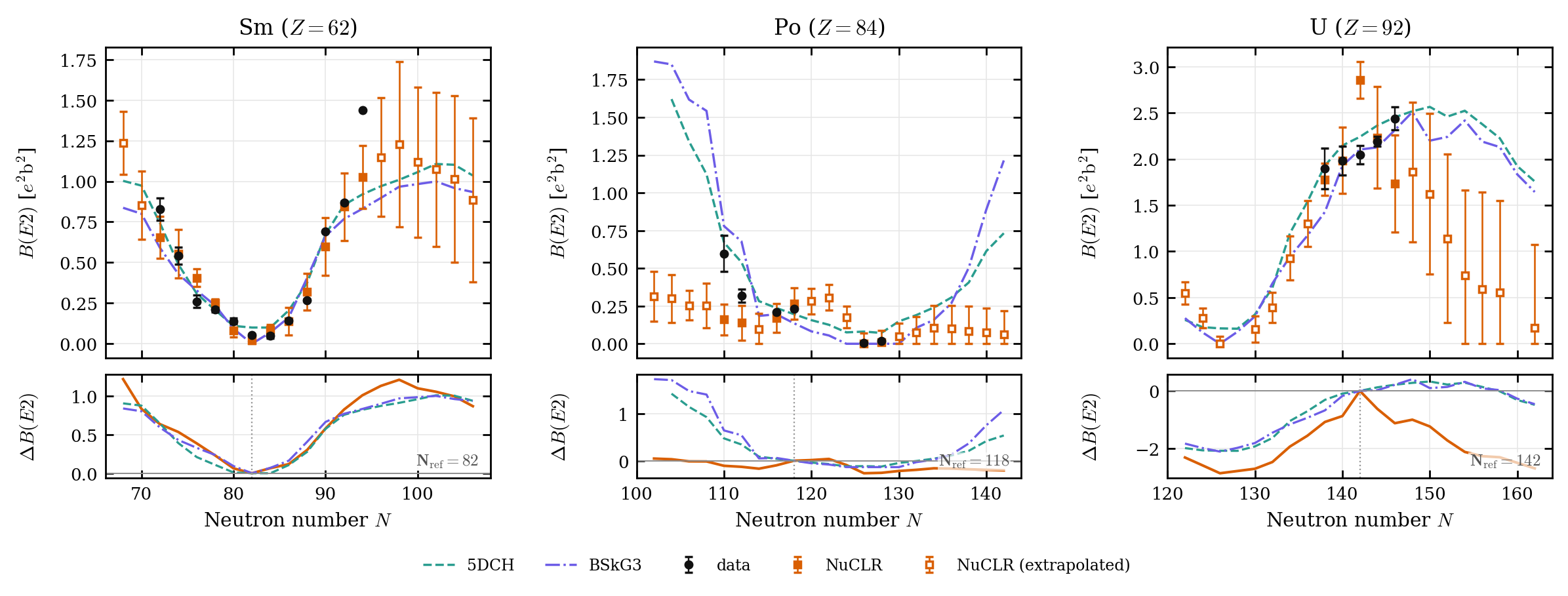}
\caption{
\BEtwo\ survey for Sm, Po, and U.
The plotting convention is the same as Fig.~\ref{fig:radii-main}, with 5DCH and BSkG3 as nuclear-model baselines.
The Po panel is shown with an expanded vertical range so the suppressed NuCLR continuation and the collective-model scale can be compared directly.
The U panel emphasizes the extrapolation across the $N=126$ region into the measured actinide rotor sequence.
Lower subpanels show $\Delta B(E2)(N)=B(E2;N)-B(E2;N_{\rm ref})$.}
\label{fig:be2-main}
\end{figure*}

\textbf{\textit{Quadrupole transitions}}~--~Because \BEtwo\ measures
quadrupole collectivity and can vary by orders of magnitude between shell
closures and deformed regions, it provides a stringent test of whether NuCLR
has learned rapidly changing trends due to many-body correlations in nuclei. Figure~\ref{fig:be2-main} shows three representative chains.

In Fig.~\ref{fig:be2-main}(left), we again show Sm, which illustrates the strong transition from near-sphericity at $N=82$ to highly-deformed shapes as one moves around that point. Particularly interesting is the behavior observed at high $N$ where data is missing, as the AI extrapolation predicts a downward trend for \BEtwo\ after $N\approx98$, also predicted by the mean-field frameworks, though with large uncertainties as the NuCLR error estimation identifies this as a region only weakly constrained purely by available data. 

In Fig.~\ref{fig:be2-main}(middle), we show polonium (Po). Experimental data are scarce, indicating low \BEtwo\ values between $110<N<130$, and an essentially spherical point at the magic number $N=126$. For these isotopes, NuCLR and the nuclear models agree well, but the picture changes dramatically in the neutron rich and deficient regions: NuCLR predicts rather flat trends as a function of $N$, whereas the nuclear models predict much steeper variations.
Analogously to the charge radii of Pb (and Hg in the Supplemental Material), this may reflect shape-coexistence effects that drive the mean-field minima toward large deformation (see \cite{Leoni:2024ivw} for a recent overview of Po results). However, the data-driven continuation does not contain any such feature, and the estimated error intervals remain mostly smaller than the discrepancy with the nuclear models. Therefore, experimental results in this region would have the power to elucidate the underlying physical mechanisms.

Finally, in Fig.~\ref{fig:be2-main}(right) we discuss the uranium (U) chain. Data are scarce, as only five experimental data points around $N=142$ exist, well reproduced by both the mean field models and NuCLR. Going to lower values of $N$, both models and NuCLR consistently predict a reduction of quadrupole collectivity toward shell closure at $N=126$. Conversely, in the neutron-rich region BSkG3 and 5DCH predict higher collectivity than NuCLR, whose predictions, however, carry large error intervals. New measurements in this region would add qualitatively new information to existing data sets.
Taken together, the error bars that NuCLR assigns to its predictions for both the neutron deficient and rich regions are interesting. 
On the neutron-deficient side, 
the learned error scale is small near the magic number \(N=126\), where nuclei with similar shell-distance features tend to have small OOF residuals. 
At large \(N\), nuclei with similar features are less reliably predicted, and the learned error scale increases accordingly.
We stress that the widths are learned from OOF residuals rather than supplied by a nuclear model, and that, 
for unmeasured nuclei, they provide structure-informed estimates of local prediction difficulty rather than intervals with validated coverage.

In the Supplemental Material, $B(E2)$ results are shown for additional representative chains, respectively, zirconium (Zr), tin (Sn), xenon (Xe), and thorium (Th). A global scan is available in the data repository.

\textbf{\textit{Conclusions}}~--~In summary, a multi-task model trained on nuclear-data tables can provide an empirical baseline for studies of nuclear structure. By sharing information across observables and evaluating targets out of fold, NuCLR determines what can be inferred from the experimental record. Comparing its predictions with nuclear models reveals where established trends are reproduced and where new measurements could distinguish theory-driven from data-driven extrapolations. The multi-task structure is essential and substantially improves 
charge-radius and \BEtwo{} \rms\ deviations relative to single-task learning.
Therefore, the charge-radius and \BEtwo\ campaigns 
extend the
task-conditioned NuCLR strategy,
and demonstrate that new
information can be incorporated coherently into a shared representation.

Indeed, this work builds on an architecture whose learned nuclear embeddings have
already been connected to physical organization in NuCLR~\cite{Kitouni:2023rct,Kitouni:2024ulw,Richardson:2025dze}.
We have not yet established how the correlations responsible for the
charge-radius and \BEtwo\ predictions emerge within that representation.
This is an important next step and could expose which observables drive
particular predictions. Moreover, motivated by the correlation between
prediction accuracy and simple measures of latent-space organization observed
in Ref.~\cite{Kitouni:2024ulw}, the geometry of the learned representation could
itself be used to predict the local error scale. This would test whether
NuCLR's internal representation contains information about the reliability of
its own predictions beyond that provided by the hand-selected
nuclear-structure and chart-location features used here.
Specific targets include the nontrivial charge-radius trends in calcium and
mercury shown in the Supplemental Material. An obvious extension is also to incorporate nuclear energy levels, for which thousands of measurements are available.

More broadly, expanding the shared
representation across nuclear observables would move NuCLR toward an AI
foundation model of the nuclear chart while strengthening its role as a
data-driven surveyor: identifying where existing data and nuclear theory
provide consistent descriptions, where they make distinguishable predictions,
and which new measurements (including from potential collider experiments \cite{STAR:2024wgy,Giacalone:2026fat}) would most effectively advance our understanding of nuclear structure.

\textbf{\textit{Acknowledgments}} -- We thank Wouter Ryssens for useful discussion, feedback on the manuscript, and for help in utilizing the BSkG3 results. We acknowledge Antoine Belley for useful discussions.  S.T. was supported by the Swiss National Science Foundation project number P5R5PT\_222350. S.T. is also supported in part by the U.S. Department of Energy (DOE) under Contracts DE-SC0023522, DE-SC0010143, and No. 89243024CSC000002 (QuantISED Program).
M.W. was supported by NSF grant PHY-2019786 (The NSF AI Institute for Artificial Intelligence and Fundamental Interactions, http://iaifi.org/).

\bibliographystyle{apsrev4-2} \bibliography{biblio}

\clearpage

\onecolumngrid

\section*{Supplemental Material}
\setcounter{figure}{0}
\renewcommand{\thefigure}{S\arabic{figure}}
\renewcommand{\theHfigure}{S\arabic{figure}}
\setcounter{table}{0}
\renewcommand{\thetable}{S\arabic{table}}
\renewcommand{\theHtable}{S\arabic{table}}

\setcounter{section}{0}
\setcounter{subsection}{0}
\renewcommand{\thesection}{S\arabic{section}}
\renewcommand{\thesubsection}{\Alph{subsection}}

\section{Additional Isotopic Chains}

Here, we present results for charge radii and quadrupole transitions of additional isotopes of particular interest.

\subsection{Charge radii}

\begin{figure}[b]
\centering
\includegraphics[width=\textwidth]{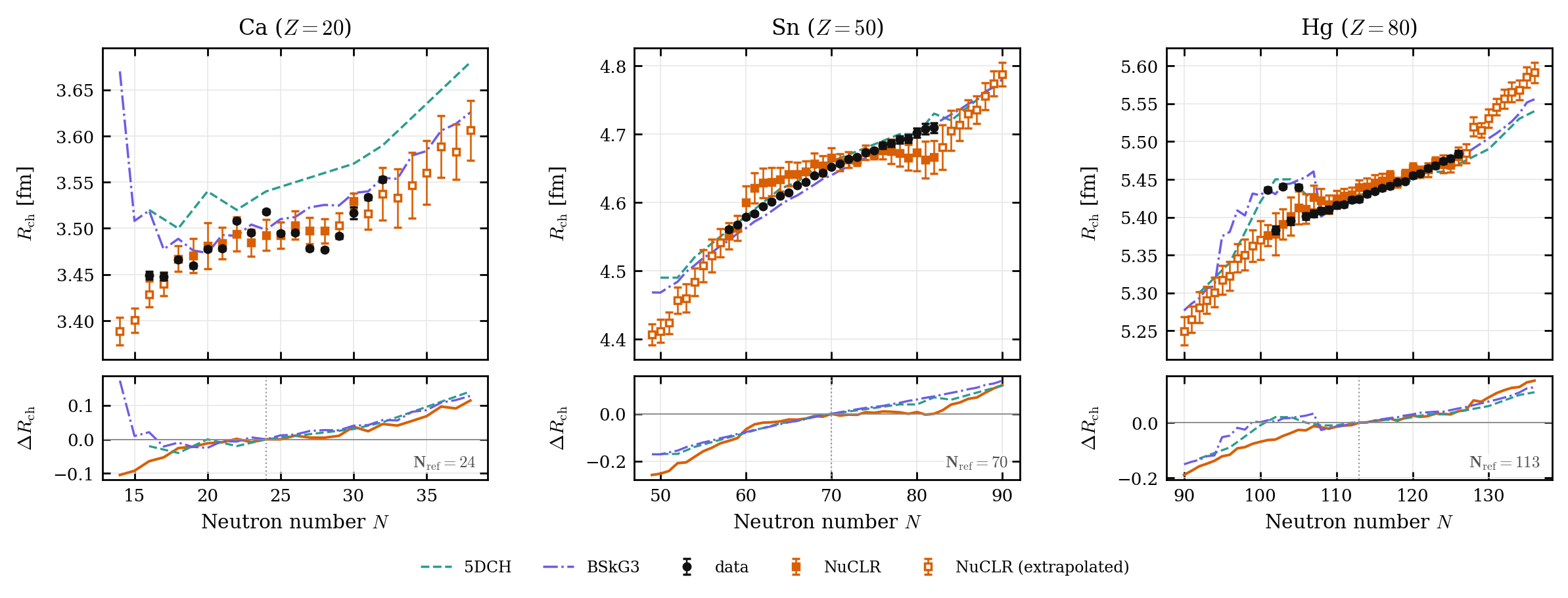}
\caption{
Additional charge-radius chains from the NuCLR ensemble: Ca, Sn, and Hg. The calcium panel includes the published $^{36,37,38}$Ca and $^{49,51,52}$Ca measurements cited in the text. The plotting and GBM-uncertainty conventions are the same as Fig.~\ref{fig:radii-main}.
}
\label{fig:appendix-radii-extra-a}
\end{figure}

In Fig.~\ref{fig:appendix-radii-extra-a}(left) we start with the chain of calcium (Ca). The experimental results, shown in the figure, rise towards $N=20$, leading to an arch-like structure with doubly magic $^{40,48}$Ca having nearly identical radii, followed by a rather steep rise for $N>28$ \cite{GarciaRuiz:2016ohj,2019NatPh..15..432M}. These features emerge from an intricate interplay of nuclear phenomena, making the calcium chain one of the most demanding benchmarks for nuclear models \cite{Reinhard:2017ugx,Perera:2021ztx,refId0,Xie:2025jmr,Naito:2026kij}, including \textit{ab initio} calculations \cite{Heinz:2024juw,10.3389/fphy.2025.1581854,Franzke:2025pvo}. 
The NuCLR intervals are too broad to determine whether the experimental arch between \(20<N<28\) can be inferred reliably from the broader nuclear data set. This region therefore provides a useful target for future studies of which observables encode the relevant structure.
NuCLR results are consistent with both data and the BSkG3 and 5DCH predictions for $N\gtrsim 20$, though for smaller $N$ NuCLR agrees better with the data than the nuclear models.

Next, in Fig.~\ref{fig:appendix-radii-extra-a}(middle) we move to the tin (Sn) chain. Much as for Ca, the chain develops around the closed proton shell with $Z=50$, which makes such isotopes particularly interesting as a benchmark for correlation effects included in phenomenological models \cite{Reinhard:2017ugx,Perera:2021ztx,PhysRevC.107.054307,Xie:2023jmb} or recent \textit{ab initio} evaluations \cite{10.3389/fphy.2025.1581854,Demol:2026glc}. 
While the nuclear models reproduce the smooth measured trend, the OOF NuCLR predictions display kinks, particularly around \(N=82\), that are not observed in the data. The broader NuCLR intervals in this region indicate that the model also identifies these predictions as less strongly supported by the patterns learned from the experimental record.

In Fig.~\ref{fig:appendix-radii-extra-a}(right) we instead focus on the chain of mercury (Hg). This chain is of great interest because it displays a large inverted odd-even staggering pattern for $100<N<105$ \cite{Marsh:2018wxs,DayGoodacre:2020byb,Goodacre:2021pny}. This is typically understood as a consequence of shape coexistence effects \cite{Heyde:2011pgw}. It challenges theoretical descriptions \cite{Sels:2019rwu,PhysRevLett.112.162701,Siciliano:2020akh,Mun:2023lfc} and has not yet been investigated in the context of \textit{ab initio} structure calculations. 
The nuclear-model calculations reproduce part of the rise toward larger radii below \(N=110\), although they do not capture the observed staggering associated with correlations. The OOF NuCLR predictions are instead smooth and likewise fail to reproduce the kink or non-monotonic behavior. Because the NuCLR intervals are narrow compared with the observed staggering, this represents a clear case in which the shared representation does not recover structure present in the measured target observable. A dedicated analysis of which additional observables encode this behavior could help elucidate the origin of the staggering and its relation to other nuclear properties.

\begin{figure}[t]
\centering
\includegraphics[width=0.5\textwidth]{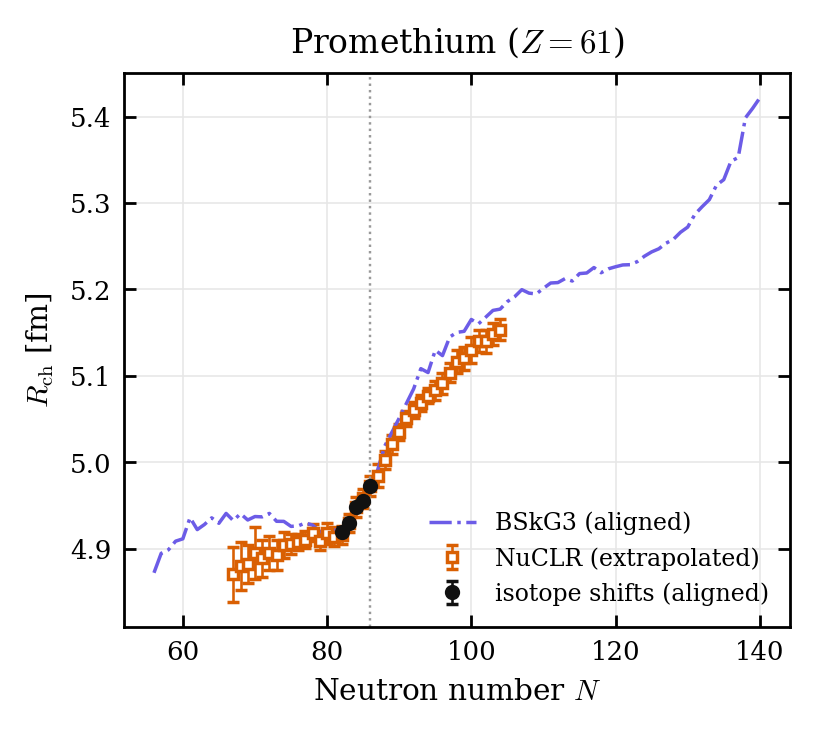}
\caption{
NuCLR predictions for the charge-radius evolution along the Pm chain. Black points show the published isotope shifts aligned to the common $^{147}$Pm NuCLR anchor; BSkG3 is aligned at the same isotope. The NuCLR bars use the GBM-uncertainty convention of the preceding figures.
}
\label{fig:promethium}
\end{figure}

Charge radii of odd-$Z$ promethium (Pm) isotopes represent potential high-impact measurements for experiments in the near future \cite{Chrysalidis:2912235}. This nucleus offers, in particular, the opportunity of studying how deformation effects are altered in the presence of an unpaired proton. We provide predictions for the charge radii along this isotopic chain in Fig.~\ref{fig:promethium}. 
The NuCLR trend agrees with both the BSkG3 prediction and the independently measured isotope shifts of Ref.~ \cite{Studer:2020hsb}.

\subsection{Quadrupole transitions}

\begin{figure}[t]
\centering
\includegraphics[width=\textwidth]{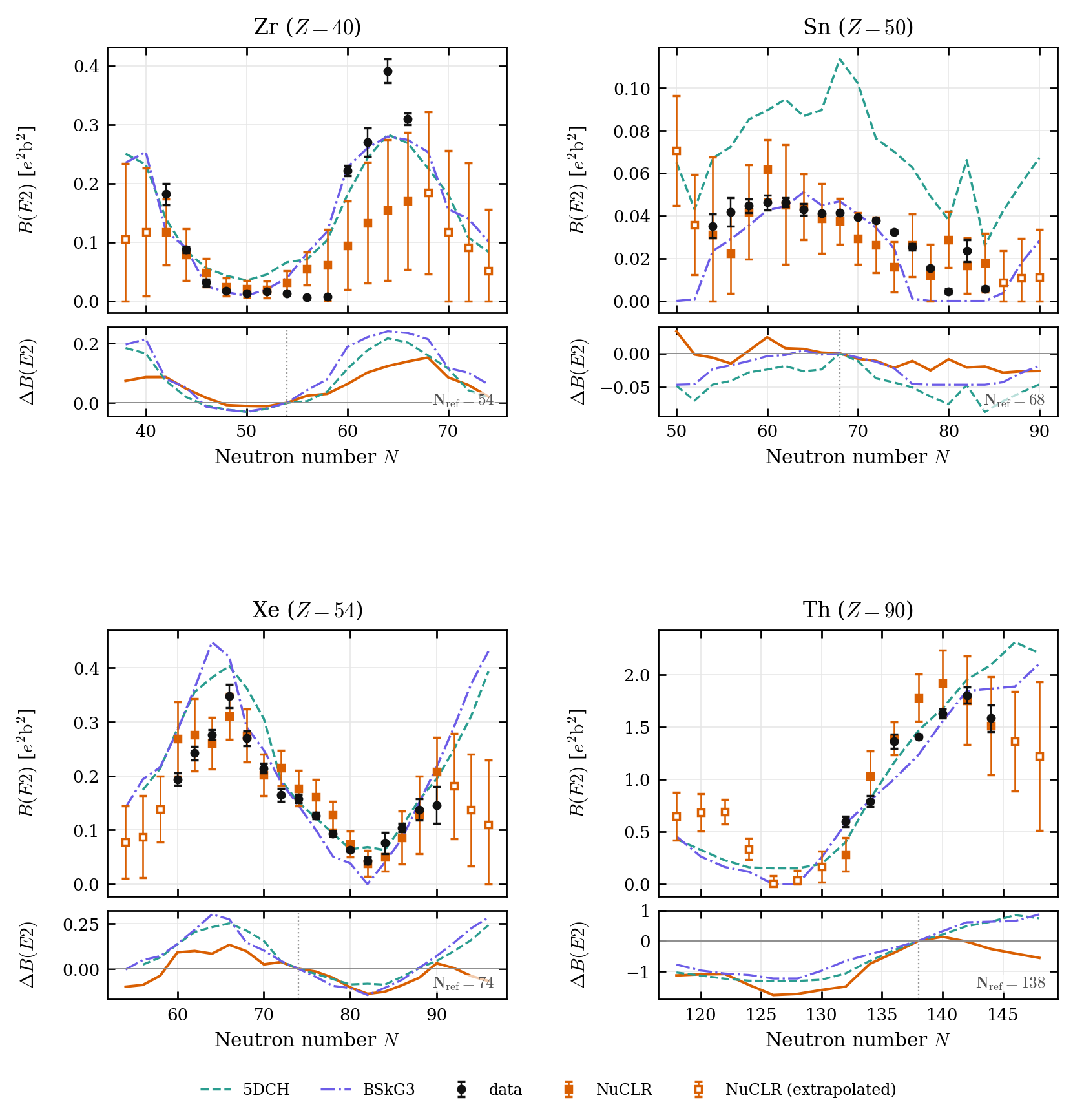}
\caption{
Additional \BEtwo\ chains from the NuCLR ensemble: Zr, Sn, Xe, and Th. The plotting and uncertainty conventions are the same as Fig.~\ref{fig:be2-main}; Sm, Po, and U are shown in the main text.
}
\label{fig:appendix-be2-extra-a}
\end{figure}

We start in Fig.~\ref{fig:appendix-be2-extra-a}(top left) with the chain of zirconium (Zr), which presents one of the most spectacular patterns of quadrupole collectivity in the nuclear chart. Starting with highly deformed isotopes around $N=40$ \cite{Hu:2024pee}, nowadays accessible in \textit{ab initio} calculations, the \BEtwo{} values then present the usual minimum around $N=50$, with $N=56$ standing out as a particularly spherical point. 
Moving beyond $N=58$, we observe a sharp increase, the most abrupt found across all nuclides. This is understood as a consequence of a quantum phase transition along the isotopic chain and the onset of strong shape coexistence effects \cite{Garcia-Ramos:2019odg,Gavrielov:2021vck,Pasqualato:2023rnt,Togashi:2016yzs}.
The intricacies of these behaviors are beyond the capabilities of conventional mean-field approaches \cite{Giacalone:2025vxa}, and make zirconium one of the most challenging testbeds for theoretical models. 
Indeed, as we see in the figure, the BSkG3 and 5DCH models only give a qualitative description of these observed trends.
The OOF NuCLR predictions likewise fail to reproduce the sharp increase in collectivity between \(N=58\) and \(N=60\), showing that this behavior cannot be inferred reliably from the observables included in the present shared representation. Determining which additional nuclear information encodes this transition would be a valuable target for a dedicated study.
In the future, it would be insightful to also study octupole deformations, given that $^{96}$Zr is known for presenting remarkably large octupole collectivity \cite{ISKRA2019396,Zhang:2021kxj,Rong:2022qez}.

Figures~\ref{fig:appendix-be2-extra-a}(top right) and \ref{fig:appendix-be2-extra-a}(bottom left) show the corresponding results for tin (Sn) and xenon (Xe), respectively. Together with Zr, these isotopic chains have the most extensive experimental coverage of \BEtwo{} values. The Sn chain exhibits a remarkably flat evolution, in contrast to the pronounced bell-shaped trend expected from simple valence-particle arguments. This nontrivial behavior is well reproduced by the nuclear models considered here and by other state-of-the-art calculations~\cite{PhysRevLett.120.152503,Togashi:2018gkq}. Xe instead displays a more conventional pattern: quadrupole collectivity increases toward the middle of the $50<N<82$ shell, decreases toward a minimum at the $N=82$ shell closure, and rises again as neutrons begin to populate the next major shell~\cite{Ilieva:2016hkt,PhysRevLett.130.052501}. All the models considered reproduce these broad trends. Most notably, the OOF NuCLR predictions distinguish the flat evolution in Sn from the shell-driven pattern in Xe, even though both chains span the same major neutron shell. This demonstrates that the shared representation captures information beyond simple proximity to magic numbers.

Finally, in Fig.~\ref{fig:appendix-be2-extra-a}(bottom right), we present thorium (Th), the actinide bridge between a shell closure and a rotor, whose deformation properties have been recently a focal point of study~\cite{Caputo:2024doz,Restrepo-Giraldo:2026blh}. Below the measured sequence, both NuCLR and the EDF models predict suppressed strengths around $N=126$ followed by a rapid return to large collectivity. 
Beyond the measured region, however, the data-driven extrapolation decreases while the nuclear-model predictions remain strongly collective. Measurements in this region would sharply discriminate between the two continuations.
In the future, it may be interesting to focus as well on predictions for $B(E3)$ values, as these actinides are expected to possess significant octupole collectivity \cite{Nomura:2013kua,Nomura:2021ivu}.

\clearpage
\section{NuCLR Architecture and Validation Details}
\label{app:method}

\subsection{Network Architecture}

\begin{figure}[t]
\centering
\includegraphics[width=\textwidth]{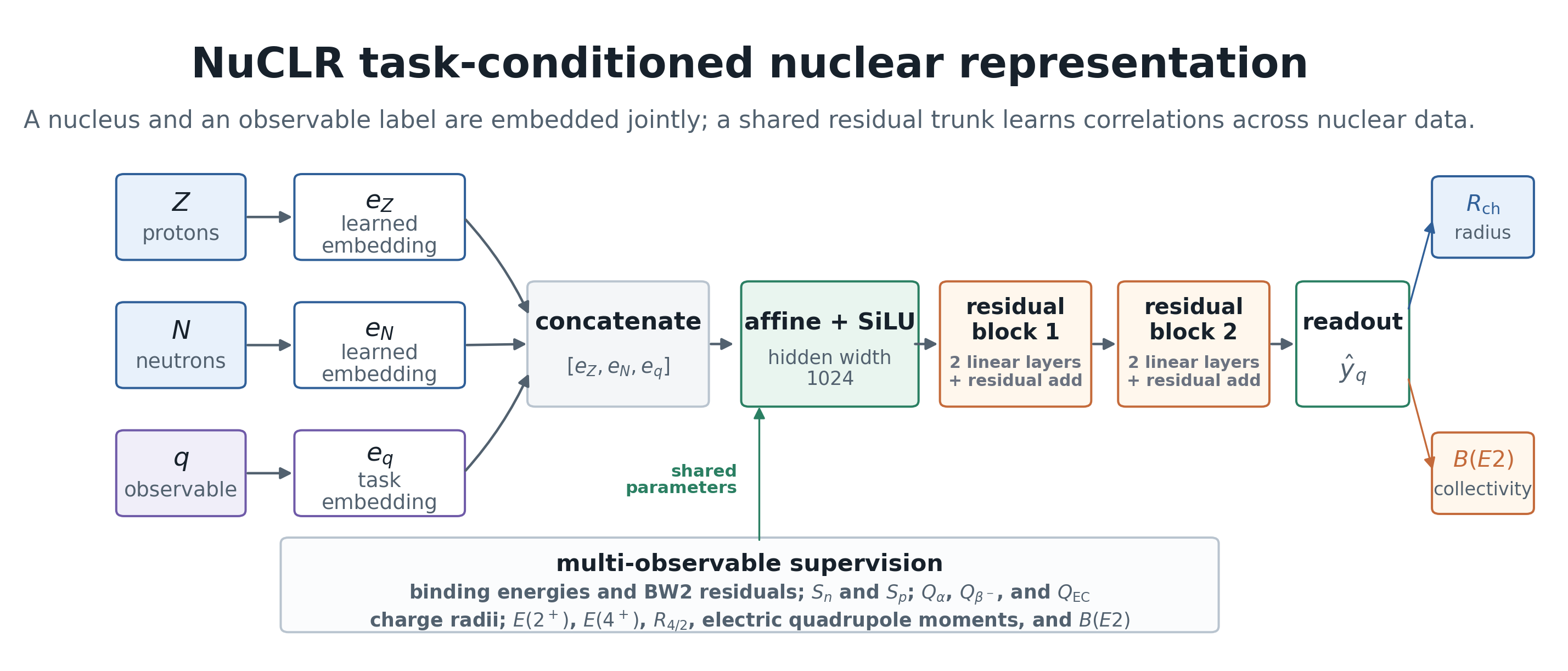}
\caption{
Schematic NuCLR architecture. The model embeds $Z$, $N$, and the observable label, learns a shared residual representation, and reads out the requested nuclear observable. The same architecture is used for the \rch\ and \BEtwo\ ensembles; the difference between the two campaigns is the pretraining and masking protocol.}
\label{fig:architecture}
\end{figure}

NuCLR treats each supervised row as a triplet $(Z,N,q)$, where $q$ is the observable label.
The proton number, neutron number, and task label are discrete tokens with learned embeddings.
The three embeddings are concatenated, passed through an affine layer with a SiLU nonlinearity, then through two residual fully connected blocks.
The campaigns use a hidden width of 1024 and the baseline residual-MLP NuCLR architecture.
Each embedding has dimension 1024, and each residual block applies two 1024-dimensional linear layers with ReLU activations followed by an identity skip connection; no dropout or normalization layer is used.
The readout produces task components, and the component associated with the requested observable is used for the loss and for the prediction.
The readout is mapped affinely to each task's target range; losses are evaluated in the tabulated task units, except for the explicit logarithmic \BEtwo\ representation described below.
Optimization uses full-batch MuAdamW with learning rate $10^{-4}$ and weight decay $10^{-2}$ on non-bias parameters. One epoch denotes one full-table optimizer update; no learning-rate schedule or early stopping is used, and the final checkpoint is retained.
The architecture is shown schematically in Fig.~\ref{fig:architecture}.

\subsection{Pretraining and Target Masking}

The radius campaign uses one shared 50000-epoch pretraining followed by multi-task finetuning.
During radius pretraining, \rch\ is excluded everywhere.
The trained tasks are the binding energy $E_b$, the BW2 binding-energy residual, the identity tasks $Z$ and $N$, one-neutron and one-proton separation energies $S_n$ and $S_p$, and the decay $Q$ values $Q_\alpha$, $Q_{\beta^-}$, and $Q_{\rm EC}$.
Radius models are then finetuned for 50000 epochs with the radius task activated while all nine support tasks remain active.
Thus, the raw binding-energy and BW2-residual supervision is retained during radius adaptation rather than frozen or removed.
The measured radius set contains 868 nuclei, divided into ten folds of 86 or 87 nuclei.
Four independently seeded models are trained for each fold, all withholding the same radius labels.

The \BEtwo\ campaign uses a more cautious protocol because several observables are direct proxies for quadrupole collectivity.
There are ten fold-specific pretrainings, one for each held-out fold, run for 50000 epochs.
The \BEtwo\ target is hidden throughout pretraining for all 433 measured nuclei: 428 even-even and five odd-odd nuclei, all with $0^+$ ground states.
The folds are stratified by \BEtwo\ magnitude and broad $Z$ region.
The related quadrupole observables $E(2_1^+)$, $B(E2;4_1^+\!\rightarrow2_1^+)$, $E(4_1^+)$, $R_{4/2}$, and the static electric quadrupole moment are also hidden on the active held-out fold.
The safe support tasks in those pretrainings are $Z$, $N$, $S_n$, $S_p$, $Q_\alpha$, the BW2 binding-energy task, and \rch.
The raw binding-energy task is not part of this campaign; its mass-related support is the BW2 residual.
The \BEtwo\ finetunings are run for 30000 epochs with four independently seeded models per fold.
Each fold withholds 42--44 measured \BEtwo\ targets.
The same cautious quadrupole mask is applied during finetuning, so the held-out target and the closest quadrupole proxies are absent together.
Positive \BEtwo\ targets are represented as $b_i=\log_{10}B_i$, while non-positive entries are excluded. Before applying the target-task weight of 2, the per-example mixed linear/log squared-error term is
\begin{equation}
 \ell_i^{B(E2)}=
 \alpha(\hat b_i-b_i)^2+
 (1-\alpha)(10^{\hat b_i}-10^{b_i})^2,
 \qquad \alpha=0.75 .
\end{equation}
Thus $\alpha$ is the weight assigned to the log-scale term, and reported predictions are transformed back to $e^2{\rm b}^2$.

Both campaigns, therefore, produce $10\times4=40$ finetuned models.
For a measured target, the central prediction is computed only from ensemble members for which that target was held out, as in Eq.~\eqref{eq:ensemble}.
For an unmeasured target, no member has seen the target value, so the full 40-member ensemble supplies the central prediction.

\subsection{Auxiliary Observables}

The auxiliary pretraining observables are written in physics notation rather than code notation here.
The BW2 task is the residual associated with a Bethe-Weizsaecker-type liquid-drop binding-energy baseline.
The separation energies are $S_n$ and $S_p$, and the decay energies are $Q_\alpha$, $Q_{\beta^-}$, and $Q_{\rm EC}$.
For the \BEtwo\ leakage-safe campaign, the cautious quadrupole proxies are $E(2_1^+)$, $E(4_1^+)$, $R_{4/2}=E(4_1^+)/E(2_1^+)$, $B(E2;4_1^+\!\rightarrow2_1^+)$, and the static electric quadrupole moment.
The auxiliary mass, separation-energy, and decay-energy labels are taken from evaluated nuclear-data tables. These tables include measured or evaluated entries as well as entries flagged as systematic estimates in AME2020; the latter are retained in the present campaigns. Accordingly, ``data driven'' does not mean that every auxiliary label is a direct measurement.

\begin{figure}[t]
\centering
\includegraphics[width=\textwidth]{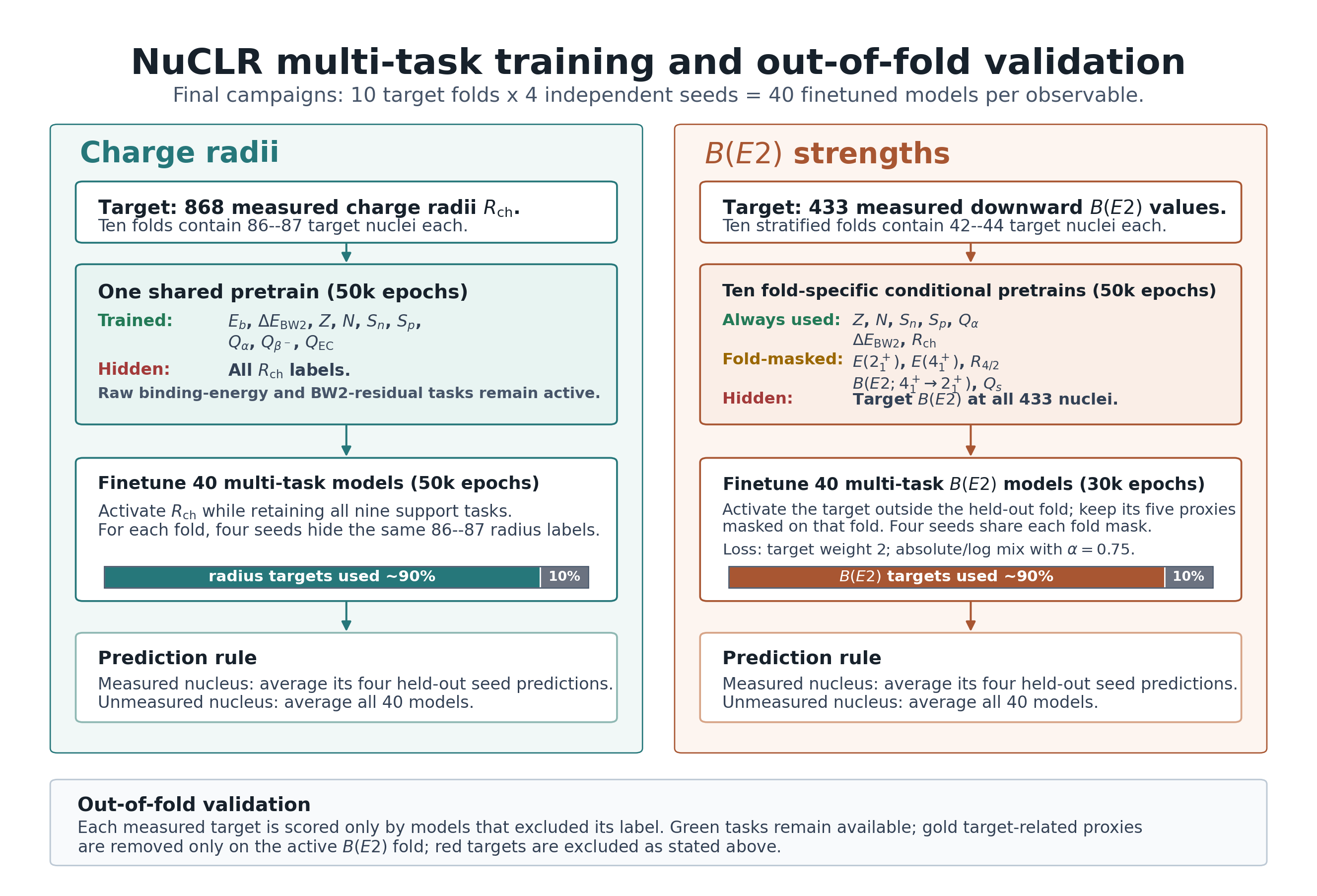}
\caption{
Complete training and masking protocol for the two NuCLR campaigns. Green tasks remain available throughout the indicated stage. Gold \BEtwo-related proxy tasks are retained away from the active validation fold but masked on that fold, together with the target, to prevent leakage. Each campaign contains ten target folds and four training seeds per fold. Measured predictions average the four models that held out that target nucleus, whereas unmeasured predictions use the full 40-model ensemble.
}
\label{fig:training-protocol}
\end{figure}

\subsection{Uncertainty Interval Model}

The purpose of the interval model is to learn where NuCLR predictions tend to be more or less accurate. It is trained only on residuals from genuinely OOF predictions. The procedure uses three disjoint groups of measured nuclei: one to train the residual-scale model, one to calibrate its overall width, and one to evaluate coverage. These roles are rotated over repeated partitions. The interval model changes only the reported widths and never the NuCLR central predictions.

The uncertainty-interval model is fitted after the NuCLR campaigns and does not retrain or alter their central predictions.
All members of ${\cal S}_i$ in Eq.~\eqref{eq:ensemble} receive equal weight.
The interval model is calibrated against the residuals of the final ensemble-averaged predictor, so no separate replica weights or ensemble-spread term are introduced. 
The calibration construction follows inductive (split) conformal regression and its normalized-nonconformity extensions, in which absolute residuals are divided by an input-dependent estimate of local prediction difficulty before a calibration quantile is applied~\cite{Papadopoulos:2002icm,Papadopoulos:2008normalized,Lei:2018distributionfree}.
Our implementation is a cross-fitted, empirically evaluated variant: the local scale is learned by gradient boosting from OOF NuCLR residual magnitudes and nuclear-structure features, and the widths are aggregated across repeated rotations as detailed below.
For a measured nucleus, define the absolute OOF residual
\begin{equation}
 r_i=|y_i-\hat y_i| .
 \label{eq:residual-scale-target}
\end{equation}
A gradient-boosted regressor $h_\phi$ then predicts one positive local residual scale,
\begin{equation}
 a_i=\exp[h_\phi(\mathbf{x}_i)] ,
 \label{eq:learned-local-scale}
\end{equation}
which is used to determine the interval widths as described below. 
Thus, \(a_i\) is not itself an error bar or a coverage interval; it is the model’s estimate of the relative residual scale at nucleus \(i\). 
The regression target is $\log[\max(r_i,\epsilon_q)]$, where
$\epsilon_q=10^{-4} \times \rms_q$ avoids the possibility of taking the logarithm of zero.

The 12-component input vector contains only nuclear-structure and chart-location quantities:
\begin{equation}
\begin{aligned}
\mathbf{x}_i=\{&Z/100,\ N/150,\ A/250,\ Z\bmod2,\ N\bmod2,\\
 &(Z\bmod2)(N\bmod2),\ |N-Z|/A,\ (N-Z)/50,\\
 &d_Z/20,\ d_N/20,\ \min(d_Z,d_N)/20,\\
 &[Z-Z_\beta(A)]/10\}.
\end{aligned}
\label{eq:residual-features}
\end{equation}
Here, $A=Z+N$, $d_Z$ and $d_N$ are distances to the nearest number in
$\{2,8,20,28,50,82,126,184\}$, and
$Z_\beta(A)=A/[1.98+0.0155A^{2/3}]$ is an empirical beta-stability line.
The selected regressor has 200 depth-2 trees, learning rate 0.05, and subsampling fraction 0.8.
The NuCLR prediction magnitude, replica spread, and distance to the nearest measured target are not inputs.
Adding the replica spread or nearest-target support features did not improve local calibration, so neither is added as a separate error term.

For each of the random five-fold partitions, three folds train $h_\phi$, one disjoint calibration fold determines the interval multiplier, and one test fold held out within that fit evaluates coverage. 
All 20 ordered calibration/test assignments are evaluated for every partition, producing 100 residual-scale fits.
For desired coverage $p$, the normalized calibration scores are
\begin{equation}
 s_j=\frac{|y_j-\hat y_j|}{a_j},\qquad j\in{\cal C},
 \label{eq:normalized-residual-score}
\end{equation}
where ${\cal C}$ denotes the calibration fold.
For $n_{\cal C}=|{\cal C}|$, let
\begin{equation}
 k_p=\left\lceil(n_{\cal C}+1)p\right\rceil,
 \qquad c_p=s_{k_p},
 \label{eq:finite-conformal-quantile}
\end{equation}
with $s_{k}$ the $k$th ordered calibration score.
The intervals and their half-width are therefore
\begin{equation}
 {\cal I}^{R_{\rm ch}}_{p,i}=
 [\hat y_i-\Delta_{p,i},\hat y_i+\Delta_{p,i}],
 \qquad
 {\cal I}^{B(E2)}_{p,i}=
 [\max\{0,\hat y_i-\Delta_{p,i}\},\hat y_i+\Delta_{p,i}],
 \qquad \Delta_{p,i}=c_p a_i .
 \label{eq:predictive-interval}
\end{equation}
The calibration factor \(c_p\) converts this relative scale into an interval with empirically tested marginal coverage on held-out measured nuclei. 
Thus $\Delta_{p,i}$ is the complete plotted half-width, not an additional component combined with the ensemble spread.

A measured nucleus receives the median half-width from the 20 fits in which it is a held-out test point; an unmeasured nucleus receives the median estimate from all 100 fits.
For this two-stage OOF construction and its repeated median aggregation, coverage is evaluated empirically rather than inferred from the finite-sample guarantee for standard split conformal prediction.
Standard conformal coverage guarantees also require exchangeability between the calibration and future examples~\cite{Lei:2018distributionfree}.
This condition is not established for extrapolations into unmeasured regions of the nuclear chart, where the target nuclei may differ systematically from the measured calibration sample.
Varying the tree depth, number of trees, and terminal-leaf constraints did not produce a configuration that consistently improved coverage, width, and local calibration for both observables.

\begin{table}[t]
\centering
\small
\setlength{\tabcolsep}{3.2pt}
\begin{tabular}{lcccccc}
\toprule
& & \multicolumn{2}{c}{Random OOF coverage}
& \multicolumn{1}{c}{68\% local score}
& \multicolumn{2}{c}{Regional coverage}\\
Observable & $n$ & 68\% & 95\% & Learned & 68\% & 95\%\\
\midrule
$R_{\rm ch}$ & 868 & 0.689(16) & 0.969(6) & 0.118(4) & 0.655(70) & 0.873(38)\\
$B(E2)$      & 433 & 0.693(22) & 0.968(9) & 0.123(5) & 0.687(92) & 0.910(26)\\
\bottomrule
\end{tabular}
\caption{Coverage and local-calibration diagnostics. Parentheses give binomial standard errors for the aggregate random-fold coverages, standard deviations over five random partitions for the local score, and standard deviations over three regional partitions for regional coverage.}
\label{tab:uncertainty-diagnostics}
\end{table}

The local score is the RMS deviation of cellwise 68\% coverage from the nominal value of 0.68 over translated $8\times8$ regions of the nuclear chart containing at least eight test nuclei; lower values indicate more uniform local calibration.
For the regional diagnostic, measured nuclei are divided into five regions by applying $k$-means clustering to standardized $(Z,N)$ coordinates. Only the residual-scale model is region-held-out; the NuCLR residuals remain the target-OOF residuals from the original ten-fold campaigns.

Randomly held-out measured nuclei show near-nominal overall coverage. Coverage is less accurate when entire contiguous regions are withheld, particularly for the nominal 95\% intervals. This degradation under regional distribution shift is why we do not interpret the widths at unmeasured nuclei as calibrated prediction intervals. For measured nuclei, filled-point bars are empirically validated OOF intervals. For unmeasured nuclei, open-point bars report the OOF residual scale learned from measured nuclei with similar structural features. 
No monotonic increase with distance from measured data is imposed. 
The 95th-percentile multiplicative error in Table~\ref{tab:metrics} is a separate tail diagnostic of the central predictions and is not part of Eqs.~\eqref{eq:learned-local-scale}--\eqref{eq:predictive-interval}.

\end{document}